# Is there any sense in antisense editing?

Yossef Neeman[1,2], Dvir Dahary[1], Erez Y. Levanon[1,3], Rotem Sorek[1,4] and Eli Eisenberg[5]

[1] *Compugen Ltd., 72 Pinchas Rosen St., Tel-Aviv 69512, Israel*

[2] *Faculty of life sciences, Bar Ilan University, Ramat Gan 52900, Israel*

[3] *Department of Pediatric Hematology-Oncology, Chaim Sheba Medical Center and Sackler School of Medicine, Tel Aviv University, Tel Aviv 52621, Israel*

[4] *Department of Human Genetics and Molecular Medicine, Sackler Faculty of Medicine, Tel Aviv University, Ramat Aviv 69978, Israel*

[5] *School of Physics and Astronomy, Raymond and Beverly Sackler Faculty of Exact Sciences, Tel Aviv University, Tel Aviv 69978, Israel*

Correspondence should be addressed to E.E. (e-mail: elieis@post.tau.ac.il).

**Teaser**: Do sense-antisense dsRNA duplexes undergo extensive A-to-I RNA editing?

Abstract:

**A number of recent studies have hypothesized that sense-antisense RNA transcript pairs create dsRNA duplexes that undergo extensive A-to-I RNA editing. Here we studied human and mouse genomic antisense regions, and found that the editing level in these areas is negligible. This observation puts in question the scope of sense-antisense duplexes formation *in-vivo*, which is the basis for a number of proposed regulatory mechanisms.**

**RNA editing**

Adenosine to inosine (A-to-I) RNA-editing is a post-transcriptional mechanism, resulting in a mature mRNA modified relative to its genomic template. A-to-I editing is mediated by members of the double-stranded RNA-specific ADAR (adenosine deaminases acting on RNA) family [1], and may change codons, create or destroy splice sites, alter RNA structure and affect RNA localization and translation rates (reviewed in [2]). RNA editing is crucial for normal life and development in both invertebrates and vertebrates [3-5], and altered editing patterns were associated with a number of pathologies. Until recently only a handful of edited human genes were documented. However, high levels of inosines are observed in mammalian transcripts, making it clear that the few editing events known were only the "tip of the iceberg" [6].

**Naturally occurring antisense RNA**

Sense and antisense transcript pairs are RNAs containing sequences that are complementary to each other. They can be transcribed in *cis*, from opposing DNA strands at the same genomic locus, or in *trans*, from distinct loci. Several independent studies have reported on widespread natural antisense transcripts (NATs). In humans, between 5% and 10% of all genes were found to have a *cis* antisense counterpart [7-9], and similar results were reported for the mouse [10], Drosophila [11], Arabidopsis [12] and rice genomes [13].

Pioneering studies in several eukaryotic systems have identified several mechanisms by which antisense transcription can regulate gene expression, including transcriptional interference, RNA masking, RNAi and RNA editing (reviewed in [14]).

Most of these mechanisms assume pairing of antisense transcripts to form double-stranded RNA (dsRNA) structures. Hypothetically, these long (average of more than 300bp) and perfect inter-molecular duplexes can serve as editing substrates. Indeed, such long and perfectly matching dsRNAs are extensively edited upon transfection to mammalian cells *in vitro* [15]. Several studies have, therefore, suggested that naturally occurring antisense dsRNA duplexes are heavily edited *in vivo*, proposing a possible general regulatory role for antisense transcripts [4,16-18]. Nevertheless, only two cases of editing in sense/antisense transcript pairs have been reported so far, none of them in mammals [19,20].

**Alu repeats as major RNA-editing substrates in human**

Recently, several independent studies have introduced bioinformatics methods for the detection of A-to-I editing sites. They found that A-to-I editing is extremely abundant in the human genome [21-24], where virtually all of the editing sites reside within Alu repeat elements. Alu elements are typically 300 nucleotides long, and account for >10% of the human genome [25]. Being so abundant in the genome, they are very likely to have a second nearby Alu repeat of reversed orientation. If such an inverted repeat exists, the two repeats can pair together to form the dsRNA hairpin structure that is then targeted by the ADARs [21-24].

While the findings of abundant Alu editing account for the observed high levels of inosines, the question whether editing of antisense transcripts plays a significant role is still open. As perfectly matching long dsRNAs are extensively edited upon transfection to mammalian cells [15], the existence of antisense editing, or lack thereof, may tell us how significant is pairing of antisense transcripts in the nucleus.

**Searching for editing sites in antisense loci**

The sequencing reaction (as well as the ribosome) recognizes inosine (I) as a guanosine (G). Therefore, the fingerprints of ADAR editing are genomically encoded adenosines that are read as guanosines in the RNA sequence. Following Ref. [26], we used alignments of 128,068 mRNA sequences to the genome in UCSC July 2003 assembly and recorded all the mismatches along them. A-to-I editing sites often occur in clusters, an edited sequence typically being edited in many close-by sites [27]. Therefore, in order to detect correct editing sites (as opposed to SNPs, sequencing and other errors), we retained only those mismatches that are part of a stretch of identical mismatches between the given RNA sequence and the genomic DNA. Applying this to all RNA sequences resulted in a vast over-representation of A-to-G mismatches compared to other common mismatches, suggesting that we indeed detect true editing sites (Table 1). We found 11,613 (~80%) clusters of three consecutive identical A-to-G mismatches, compared to only 968 such clusters of G-to-A mismatches (~7%). This means that roughly 10,600 (>90%) of the detected A-to-G mismatch clusters are probably a result of A-to-I editing events [26]. The specificity improves as we increase the number of identical mismatches in the cluster: Requiring stretches of five consecutive identical mismatches results in 96% A-to-G mismatch clusters (Table 1).

To test whether antisense RNAs are significant substrates for RNA editing, we combined our algorithm for identifying antisense regions [7] with the above approach for detection of editing sites [26]. Using the Antisensor algorithm [7] (see textbox), we found pairs of overlapping transcriptional units on opposite DNA strands. This approach yielded 9,502 genes, which are predicted to have a *cis* antisense counterpart. These genomic loci, in which sense and antisense transcripts are predicted to overlap, cover a total of ~4.3 million base-pairs and are supported by 16,344 RNA sequences (GenBank version GB139). The average overlap was 316bp and the median was 168bp. We then looked for traces of A-to-I editing in sequences transcribed from antisense loci. This resulted in 4,307 RNA sequences that include at least one mismatch within an antisense region.

**Antisense regions are not extensively edited**

Focusing on antisense regions only, we detected a pattern similar to that detected for the entire RNA sequences (Table 1). For clusters of three consecutive mismatches ~78% are of the A-to-G type (~80% in the entire RNA sequences), and for clusters of five consecutive mismatches ~95% are A-to-G mismatches (~96% in the entire RNAs). However, one must take into account the occurance of editing sites in the antisense loci due to intra-molecular Alu-Alu dsRNAs. Indeed, a recent study by Athanasiadis et al. [21] have reported that only 1% of editing in Alu elements could be attributed to inter-molecular dsRNAs, with the rest probably guided by intra-molecular Alu-Alu dsRNA. We therefore filtered out all Alu repeats from the dataset using Repeatmasker (<http://www.repeatmasker.org/>) results downloaded from the UCSC database. These Alu repeats comprise merely 7% of the antisense genomic loci (~320,000 base-pairs). Yet, excluding this small fraction of the antisense regions, the overrepresentation of A-to-G mismatch clusters over other common mismatches (Figure 1) virtually vanished, suggesting that antisense transcripts, apart from the Alu regions within them, are not extensively edited. Our results do show a slight preference of A-to-G mismatches over other mismatches (18 events of A-to-G mismatch clusters vs. 8 events of T-to-C, see Table 1). Some of theses mismatches might attest for RNA editing events, either due to intra-molecular paring other then Alu-Alu pairing, or as an outcome of antisense transcripts pairing. At any rate, these few examples are negligible compared to the abundance of the global RNA editing phenomenon in human.

To rule out the possibility that RNAs having antisense counterparts have different features than RNAs not having such counterpart sequence, we repeated the analysis for the subset of RNAs that are part of antisense pairs, and searched for differences between their overlapping and the non-overlapping regions. The results were essentially identical: after filtering out Alu elements, both regions have the same mismatch distribution exhibiting no significant overrepresentation of A-to-G mismatch clusters. For example, among the mismatch clusters of length 3, 46% and 38% are A-to-G in overlapping and non-overlapping regions, respectively, with a very low overall mismatches rate (see Supplementary Table 1).

The widespread natural antisense transcripts (NATs) phenomenon is not human specific. ~2,400 sense-antisense gene pairs have been identified in the mouse genome [10], 1,027 in the Drosophila genome and a similar number is predicted to exist in the C. elegans genome [14]. Although it was expected that similar levels of editing would be observed for all mammals, two recent studies reported that the editing levels in human are at least an order of magnitude higher than that of mouse, with most human editing sites residing within Alu elements [22,26]. It is thus possible that antisense editing in human is masked by the huge amount of Alu editing, but will be observable in other organisms. We therefore applied the same search algorithm to the mouse data. Here too, we found no preference for A-to-G mismatches over other common mismatches in antisense region (see Supplementary Table 2).

**Conclusions**

It had been shown that long, perfectly matching dsRNAs are extensively edited upon transfection to mammalian cells [15]. However, Athanasiadis et al have recently reported that only 1% of editing in Alus could be attributed to inter-molecular dsRNA, suggesting that antisense pairing does not lead to significant editing [21]. Here we have conducted a systematic search throughout human and mouse antisense loci, looking for traces of A-to-I RNA editing. While our results cannot exclude the possibility of some antisense genes being modified by editing, we found no evidence for significant RNA editing within antisense regions. These results might lead to the conclusion that inter-molecular sense-antisense RNA pairings do not normally occur after transcription in the nucleus. Alternatively, pairing might actually occur within the cell, but the resulting duplexes, edited or unedited, are either retained in the nucleus [18] or degraded by RNAi or other mechanism, and are thus not represented in expressed sequence data. The role of the abundant antisense transcripts in gene expression regulation is, therefore, yet elusive.

**Acknowledgements**

R.S. was supported by a fellowship from the Clore Israel Foundation. E.E. was supported by an Alon Fellowship at Tel Aviv University.

# Figures and tables:

**Figure 1**.

**Distributions of instances of consecutive three identical mismatches, for the different mismatch types**. a. Genomic mismatches in all RNA sequences (UCSC July 2003 assembly); b. Genomic mismatches in antisense regions; c. Genomic mismatches in antisense regions without Alu repeats. A significant preference of A-to-G mismatches is observed both for all RNAs and for antisense regions. However, after filtering out Alu repeats regions, there is no significant preference for A-to-G over other mismatches, suggesting that intra-molecular pairing of Alu repeats is responsible for the observed editing signal.

a.

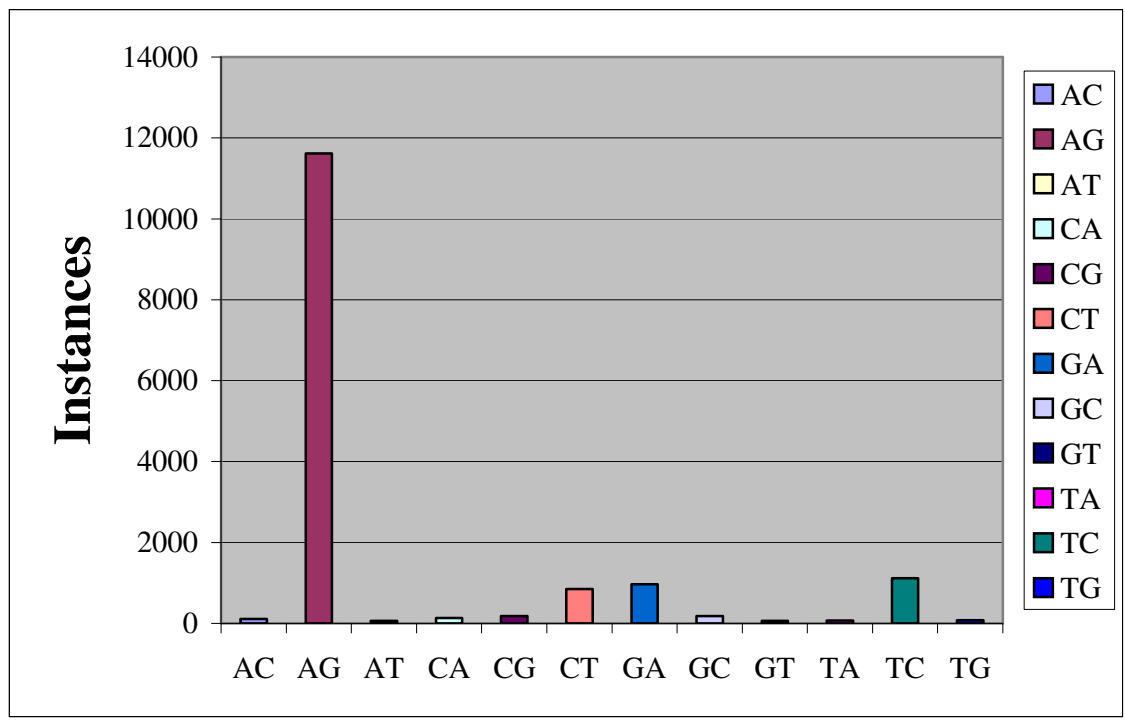

b.

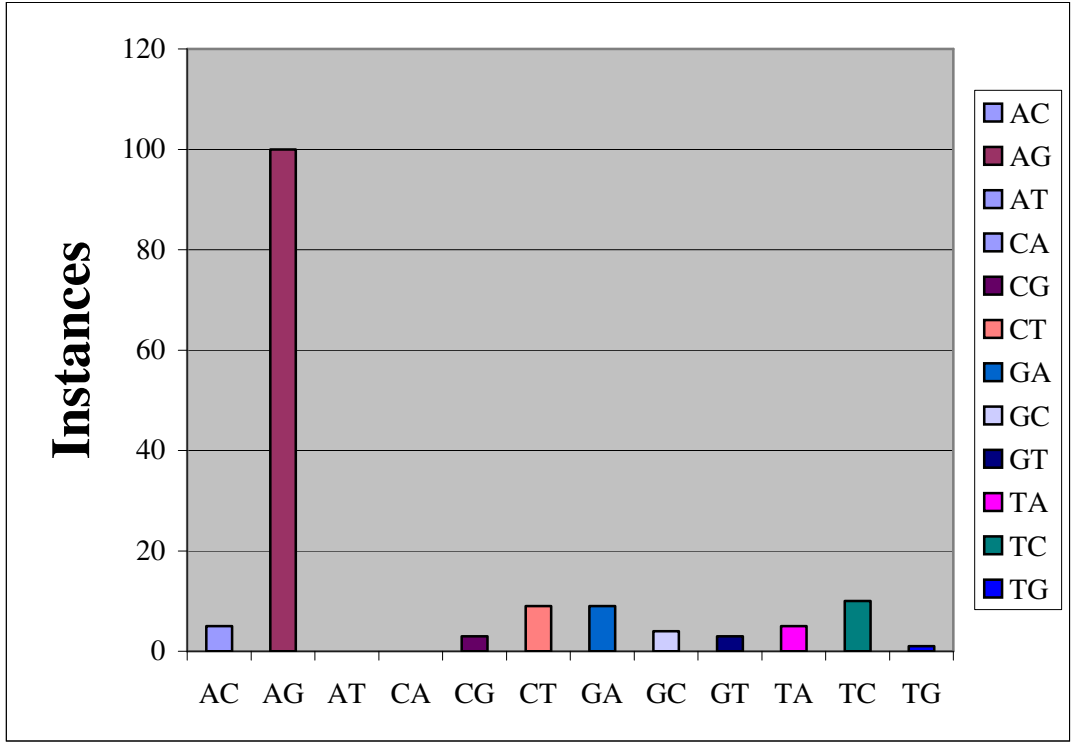

c.

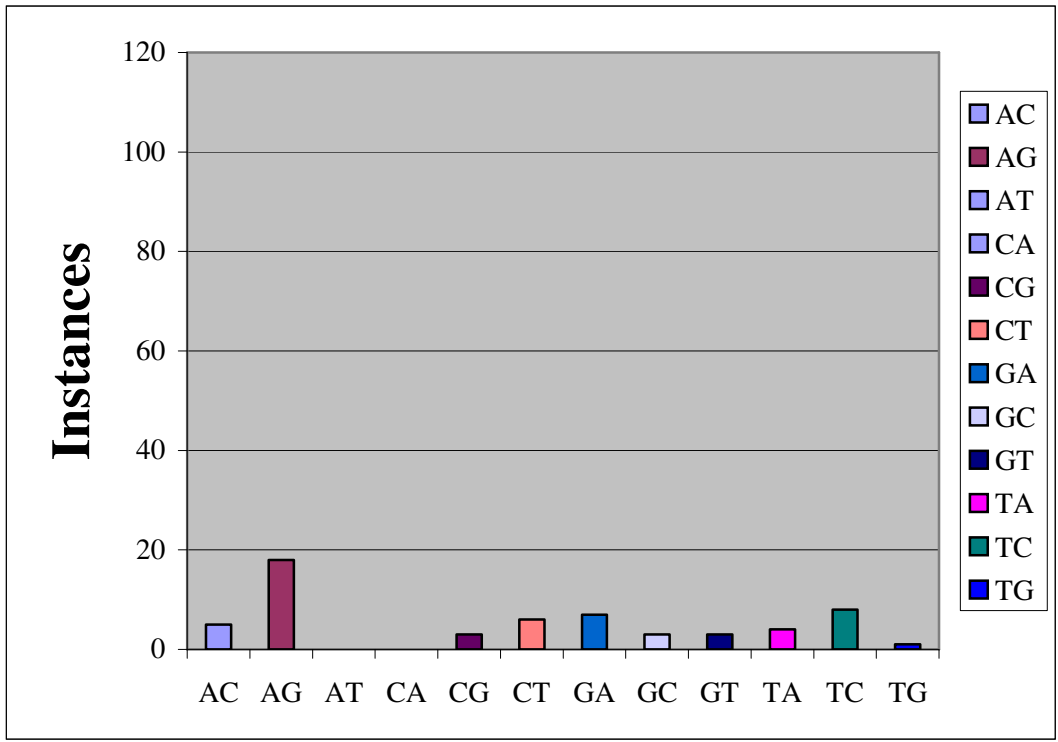

**Figure 2.**

**Editing within antisense region is restricted to Alu repeats.** A typical example of RNA-editing within antisense region, attributed to Alu pairing. Guanine nucleotide binding protein beta polypeptide 1-like (GNB1L) has an overlapping antisense region with T-box 1 (TBX1). The genomic overlap region includes an Alu repeat. A cluster of A-to-G mismatches (highlighted), indicative of A-to-I editing, is found only in the Alu region, while the rest of the antisense region is not edited at all.

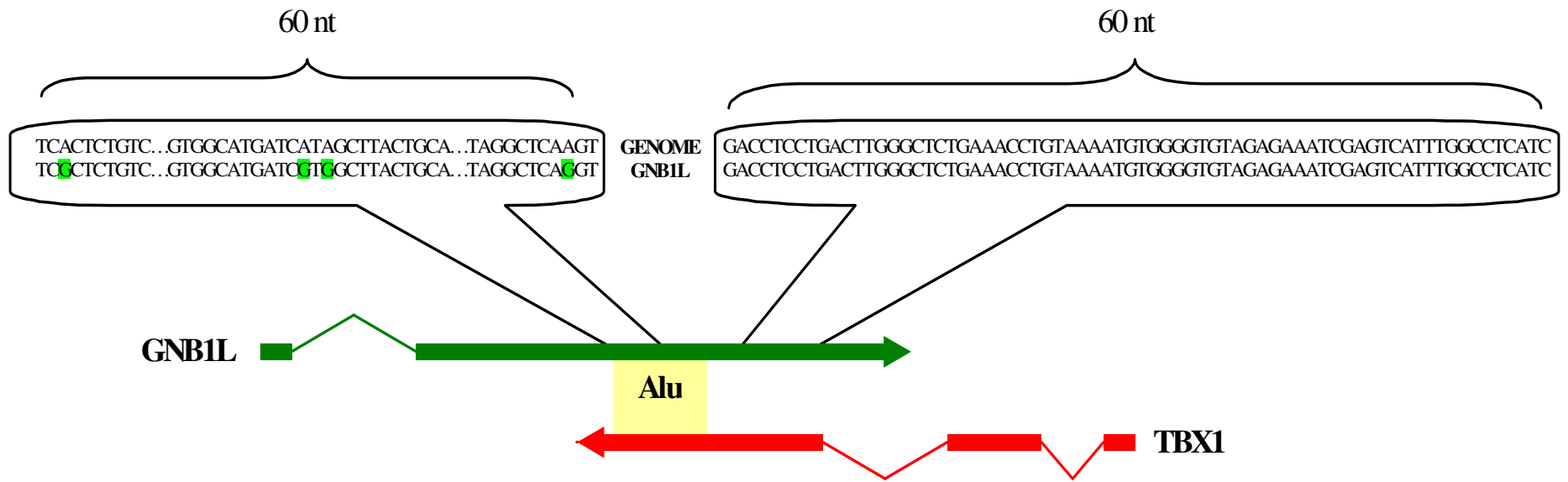

# Textbox:

**Antisensor** - an algorithm that assigns the correct DNA strand for each expressed sequence. The two main sources of information used are the splice junction consensus sequences (introns begin with GT and end with AG in most known introns), and the polyA tail at the 3` end of the transcripts. Cluster of sequences coming from a given genomic locus are then separated according to the strand the sequences are transcribed from. If two distinct reliable sequence clusters are formed, one deduces the existence of antisense transcription.

In the figure, thin bars represent the two strands of DNA; thicker bars stand for expressed sequences, where lines connect the different exons. Splice junction consensus sequences and polyA tail sequences are indicated. Based on these sources of information, the expressed sequences are separated into two clusters, colored according to the DNA strand they are transcribed from.

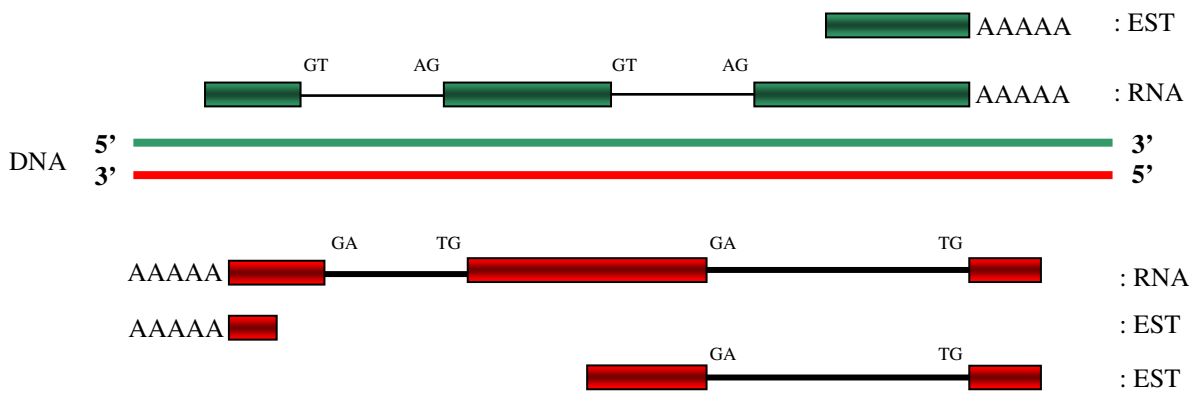

**Table 1.** Mismatch clusters in all human RNAs and Antisense regions[a]

|  |  | All RNA (94858) | | Antisense regions only | | Antisense regions without repeats | |
|---|---|---|---|---|---|---|---|
|  | Mismatch type | Number | Percent | Number | Percent | Number | Percent |
| Cluster of 1 | A to G | 102832 | 39% | 1731 | 34% | 1286 | 29% |
|  | G to A | 52488 | 20% | 1029 | 20% | 945 | 21% |
|  | T to C | 58083 | 22% | 1317 | 25% | 1191 | 27% |
|  | C to T | 52195 | 19% | 1076 | 21% | 1002 | 23% |
| Cluster of 3 | A to G | 11613 | 80% | 100 | 78% | 18 | 46% |
|  | G to A | 968 | 6% | 9 | 7% | 7 | 18% |
|  | T to C | 1115 | 8% | 10 | 8% | 8 | 21% |
|  | C to T | 853 | 6% | 9 | 7% | 6 | 15% |
| Cluster of 5 | A to G | 4926 | 96% | 37 | 95% | 3 | 60% |
|  | G to A | 48 | 1% | 0 | 0% | 0 | 0 |
|  | T to C | 71 | 1.5% | 2 | 5% | 2 | 40% |
|  | C to T | 74 | 1.5% | 0 | 0% | 0 | 0 |

[a]The number of a single or a cluster of consecutive mismatches, for the four most common mismatches. In the complete sequence of the RNAs, as well as in antisense regions only, there is a preference for A-to-G mismatches over all other mismatches. When filtering out Alu repeats in antisense regions (third column), the distribution of A-to-G over other mismatches sharply drops, and the number of mismatches becomes negligible.

**Supplementary table 1.** Human mismatch clusters distribution for antisense and non-antisense regions[a]

|  | Mismatch type | Antisense regions | | Antisense regions excluding Alu repeats | | Non antisense regions | | Non antisense regions excluding Alu repeats | |
|---|---|---|---|---|---|---|---|---|---|
|  |  | Number | Percent | Number | Percent | Number | Percent | Number | Percent |
| Cluster of 1 | A to G | 1731 | 34% | 1286 | 29% | 3382 | 37% | 2113 | 29% |
|  | G to A | 1029 | 20% | 945 | 21% | 1816 | 20% | 1643 | 23% |
|  | T to C | 1317 | 25% | 1191 | 27% | 2007 | 22% | 1783 | 24% |
|  | C to T | 1076 | 21% | 1002 | 23% | 1980 | 21% | 1751 | 24% |
| Cluster of 3 | A to G | 100 | 78% | 18 | 46% | 340 | 75% | 56 | 38% |
|  | G to A | 9 | 7% | 7 | 18% | 37 | 8% | 32 | 21% |
|  | T to C | 10 | 8% | 8 | 21% | 37 | 8% | 30 | 20% |
|  | C to T | 9 | 7% | 6 | 15% | 39 | 9% | 31 | 21% |
| Cluster of 5 | A to G | 37 | 95% | 3 | 60% | 135 | 96% | 8 | 73% |
|  | G to A | 0 | 0 | 0 | 0 | 1 | 1% | 1 | 9% |
|  | T to C | 2 | 5% | 2 | 40% | 2 | 1% | 1 | 9% |
|  | C to T | 0 | 0 | 0 | 0 | 3 | 2% | 1 | 9% |

[a] The table presents the mismatch clusters distributions for RNA sequences which have an antisense counterpart, comparing their antisense region to other parts of the RNA sequence. Antisense sequences show the same mismatch distribution in the antisense regions and in non-antisense regions. Before filtering Alu repeat elements there is a preference for A-to-G mismatches over other common mismatches, while after filtering Alu repeats the preference is gone.

**Supplementary table 2.** Mismatch clusters in all mouse RNAs and Antisense regions[a]

|  | Mismatche type | All RNA (94858) | | Antisense regions only | | Antisense regions without B1 repeats | |
|---|---|---|---|---|---|---|---|
|  |  | Number | Percent | Number | Percent | Number | Percent |
| Cluster of 1 | A to G | 38910 | 27% | 1030 | 26% | 952 | 25% |
|  | G to A | 35876 | 25% | 1045 | 26% | 993 | 26% |
|  | T to C | 31630 | 22% | 850 | 21% | 811 | 21% |
|  | C to T | 36750 | 26% | 1106 | 27% | 1038 | 28% |
| Cluster of 3 | A to G | 1112 | 31% | 11 | 20% | 4 | 8% |
|  | G to A | 748 | 21% | 13 | 22% | 11 | 23% |
|  | T to C | 742 | 20% | 10 | 17% | 10 | 21% |
|  | C to T | 1033 | 28% | 24 | 41% | 23 | 28% |
| Cluster of 5 | A to G | 181 | 46% | 2 | 33% | 0 | 0 |
|  | G to A | 64 | 16% | 0 | 0 | 0 | 0 |
|  | T to C | 49 | 13% | 0 | 0 | 0 | 0 |
|  | C to T | 98 | 25% | 4 | 67% | 4 | 100% |

[a]The number of a single or a cluster of consecutive mismatches, for the most common mismatches. While in all mouse RNAs there is a slight preference for A-to-G mismatches over other mismatches, when focusing on antisense regions there is no preference.